\documentclass[titlepage,letterpaper,12pt]{article}
\usepackage[dvips]{graphicx}

\begin{document}

\title{Bounded solutions of fermions in the background of mixed vector-scalar
inversely linear potentials}
\date{}
\author{Antonio S. de Castro \\
\\
UNESP - Campus de Guaratinguet\'{a}\\
Departamento de F\'{\i}sica e Qu\'{\i}mica\\
Caixa Postal 205\\
12516-410 Guaratinguet\'{a} SP - Brasil\\
\\
E-mail address: castro@feg.unesp.br (A.S. de Castro)}
\date{}
\maketitle

\begin{abstract}
The problem of a fermion subject to a general mixing of vector and scalar
potentials in a two-dimensional world is mapped into a Sturm-Liouville
problem. Isolated bounded solutions are also searched. For the specific case
of an inversely linear potential, which gives rise to an effective Kratzer
potential in the Sturm-Liouville problem, exact bounded solutions are found
in closed form. The case of a pure scalar potential with their isolated
zero-energy solutions, already analyzed in a previous work, is obtained as a
particular case. The behaviour of the upper and lower components of the
Dirac spinor is discussed in detail and some unusual results are revealed.
The nonrelativistic limit of our results adds a new support to the
conclusion that even-parity solutions to the nonrelativistic one-dimensional
hydrogen atom do not exist.
\end{abstract}

\section{Introduction}

The problem of a particle subject to an inversely linear potential in one
spatial dimension ($\sim |x|^{-1}$), known as the one-dimensional hydrogen
atom, has received considerable attention in the literature (for a rather
comprehensive list of references, see \cite{xia}). This problem presents
some conundrums regarding the parities of the bound-state solutions and the
most perplexing is that one regarding the ground state. Loudon \cite{lou}
claims that the nonrelativistic Schr\"{o}dinger equation provides a
ground-state solution with infinite eigenenergy and a related eigenfunction
given by a delta function centered about the origin. This problem was also
analyzed with the Klein-Gordon equation and there it was revealed a finite
eigenenergy and an exponentially decreasing eigenfunction \cite{spe}. By
using the technique of continuous dimensionality the problem was approached
with the Schr\"{o}dinger, Klein-Gordon and Dirac equations \cite{mos}. The
conclusion in this last work reinforces the claim of Loudon. Furthermore,
the author of Ref. \cite{mos} concludes that the Klein-Gordon equation
provides unacceptable solutions while the Dirac equation, with the
interacting potential considered as a time component of a vector, has no
bounded solutions at all. On the other hand, in a more recent work \cite{xia}
the authors use connection conditions for the eigenfunctions and their first
derivatives across the singularity of the potential, and conclude that only
the odd-parity solutions of the Schr\"{o}dinger equation survive. The
relativistic problem of a fermion in an inversely linear potential was also
sketched for a Lorentz scalar potential in the Dirac equation \cite{ho}, but
the analysis is incomplete. In a recent work \cite{asc6} it was shown that
the problem of a fermion under the influence of a general scalar potential
for nonzero eigenenergies can be mapped into a Sturm-Liouville problem.
Next, the key conditions for the existence of bound-state solutions were
settled for power-law potentials, and the possible zero-mode solutions were
shown to conform with the ultrarelativistic limit of the theory. In
addition, the solution for an inversely linear potential was obtained in
closed form. The effective potential resulting from the mapping has the form
of the Kratzer potential \cite{kra}. It is noticeable that this problem has
an infinite number of acceptable bounded solutions, nevertheless it has no
nonrelativistic limit for small quantum numbers. It was also shown that in
the regime of strong coupling additional zero-energy solutions can be
obtained as a limit case of nonzero-energy solutions. The ideas of
supersymmetry had already been used to explore the two-dimensional Dirac
equation with a scalar potential \cite{coo}-\cite{nog}, nevertheless the
power-law potential has been excluded of such discussions.

The Coulomb potential of a point electric charge in a 1+1 dimension,
considered as the time component of a Lorentz vector, is linear ($\sim
\left| x\right| $) and so it provides a constant electric field always
pointing to, or from, the point charge. This problem is related to the
confinement of fermions in the Schwinger and in the massive Schwinger models
\cite{col1}-\cite{col2}, and in the Thirring-Schwinger model \cite{fro}. It
is frustrating that, due to the tunneling effect (Klein\'{}s paradox), there
are no bound states for this kind of potential regardless of the strength of
the potential \cite{cap}-\cite{gal}. The linear potential, considered as a
Lorentz scalar, is also related to the quarkonium model in one-plus-one
dimensions \cite{hoo}-\cite{kog}. Recently it was incorrectly concluded that
even in this case there is solely one bound state \cite{bha}. Later, the
proper solutions for this last problem were found \cite{cas1}-\cite{hil}.
However, it is well known from the quarkonium phenomenology in the real 3+1
dimensional world that the best fit for meson spectroscopy is found for a
convenient mixture of vector and scalar potentials put by hand in the
equations (see, \textit{e.g.}, \cite{luc}). The same can be said about the
treatment of the nuclear phenomena describing the influence of the nuclear
medium on the nucleons \cite{ser}-\cite{nosso}. The mixed vector-scalar
potential has also been analyzed in 1+1 dimensions for a linear potential
\cite{cas2} as well as for a general potential which goes to infinity as $%
|x|\rightarrow \infty $ \cite{ntd}. In both of those last references it has
been concluded that there is confinement if the scalar coupling is of
sufficient intensity compared to the vector coupling.

Motived by the success found in Ref. \cite{asc6} we re-examine the
two-dimensional problem of a fermion in the background of an inversely
linear potential by considering a convenient mixing of vector and scalar
Lorentz structures, that is to say $|V_{s}|\geq $ $|V_{v}|$. The problem is
mapped into an exactly solvable Sturm-Liouville problem of a
Schr\"{o}dinger-like equation with an effective Kratzer potential. The case
of a pure scalar potential with their isolated zero-energy solutions,
already analyzed  \cite{asc6}, is obtained as a particular case. Those
ultrarelativistic zero-eigenmodes emerge despite the well-defined parity of
the potential. Our results for $V_{v}=V_{s}$ give new support to the
conclusion that even-parity solutions to the nonrelativistic one-dimensional
hydrogen atom do not exist.

\section{The Dirac equation with mixed vector-scalar potentials in a 1+1
dimension}

In the presence of time-independent vector and scalar potentials the 1+1
dimensional time-independent Dirac equation for a fermion of rest mass $m$
reads

\begin{equation}
\mathcal{H}\Psi =E\Psi  \label{1a}
\end{equation}

\begin{equation}
\mathcal{H}=c\alpha p+\beta \left( mc^{2}+V_{s}\right) +V_{v}  \label{1b}
\end{equation}

\noindent where $E$ is the energy of the fermion, $c$ is the velocity of
light and $p$ is the momentum operator. $\alpha $ and $\beta $ are Hermitian
square matrices satisfying the relations $\alpha ^{2}=\beta ^{2}=1$, $%
\left\{ \alpha ,\beta \right\} =0$. From the last two relations it follows
that both $\alpha $ and $\beta $ are traceless and have eigenvalues equal to
$\pm $1, so that one can conclude that $\alpha $ and $\beta $ are
even-dimensional matrices. One can choose the 2$\times $2 Pauli matrices
satisfying the same algebra as $\alpha $ and $\beta $, resulting in a
2-component spinor $\Psi $. The vector and scalar potentials are given by $%
V_{v}$ and $V_{s}$ respectively. The positive definite function $|\Psi
|^{2}=\Psi ^{\dagger }\Psi $, satisfying a continuity equation, is
interpreted as a position probability density and its norm is a constant of
motion. This interpretation is completely satisfactory for single-particle
states \cite{tha}. We use $\alpha =\sigma _{1}$ and $\beta =\sigma _{3}$.
The subscripts for the terms of potential denote their properties under a
Lorentz transformation: $v$ for the time component of the 2-vector potential
and $s$ for the scalar term, respectively. It is worth to note that the
Dirac equation is covariant under $x\rightarrow -x$ if $V_{v}(x)$ and $%
V_{s}(x)$ remain the same. This is because the parity operator $P=\exp
(i\eta )P_{0}\sigma _{3}$, where $\eta $ is a constant phase and $P_{0}$
changes $x$ into $-x$, changes sign of $\alpha $ but not of $\beta $.

Provided that the spinor is written in terms of the upper and the lower
components, $\Psi _{+}$ and $\Psi _{-}$ respectively, \noindent the Dirac
equation decomposes into:
\begin{equation}
i\hbar c\Psi _{\pm }^{\prime }=\left[ V_{v}-E\mp \left( mc^{2}+V_{s}\right)
\right] \Psi _{\mp }  \label{1c}
\end{equation}

\noindent where the prime denotes differentiation with respect to $x$. In
terms of $\Psi _{+}$ and $\Psi _{-}$ the spinor is normalized as $%
\int_{-\infty }^{+\infty }dx\left( |\Psi _{+}|^{2}+|\Psi _{-}|^{2}\right) =1$%
, so that $\Psi _{+}$ and $\Psi _{-}$ are square integrable functions. It is
clear from the pair of coupled first-order differential equations (\ref{1c})
that both $\Psi _{+}$ and $\Psi _{-}$ have opposite parities if the Dirac
equation is covariant under $x\rightarrow -x$.

In the nonrelativistic approximation (potential energies small compared to $%
mc^{2}$ and $E\simeq mc^{2}$) Eq. (\ref{1c}) becomes

\begin{equation}
\Psi _{-}=\frac{p}{2mc}\Psi _{+}  \label{1d}
\end{equation}

\begin{equation}
\left( -\frac{\hbar ^{2}}{2m}\frac{d^{2}}{dx^{2}}+V_{v}+V_{s}\right) \Psi
_{+}=\left( E-mc^{2}\right) \Psi _{+}  \label{1e}
\end{equation}

\noindent Eq. (\ref{1d}) shows that $\Psi _{-}$ is of order $v/c<<1$
relative to $\Psi _{+}$ and Eq. (\ref{1e}) shows that $\Psi _{+}$ obeys the
Schr\"{o}dinger equation with binding energy equal to $E-mc^{2}$, and
without distinguishing the contributions of vector and scalar potentials.

It is remarkable that the Dirac equation with a scalar potential, or a
vector potential contaminated with some scalar coupling, is not invariant
under $V\rightarrow V+const.$, this is so because only the vector potential
couples to the positive-energies in the same way it couples to the
negative-ones, whereas the scalar potential couples to the mass of the
fermion. Therefore, if there is any scalar coupling the absolute values of
the energy will have physical significance and the freedom to choose a
zero-energy will be lost. It is well known that a confining potential in the
nonrelativistic approach is not confining in the relativistic approach when
it is considered as a Lorentz vector. It is surprising that relativistic
confining potentials may result in nonconfinement in the nonrelativistic
approach. This last phenomenon is a consequence of the fact that vector and
scalar potentials couple differently in the Dirac equation whereas there is
no such distinction among them in the Schr\"{o}dinger equation. This
observation permit us to conclude that even a ``repulsive'' potential can be
a confining potential. The case $V_{v}=-V_{s}$ presents bounded solutions in
the relativistic approach, although it reduces to the free-particle problem
in the nonrelativistic limit. The attractive vector potential for a fermion
is, of course, repulsive for its corresponding antifermion, and vice versa.
However, the attractive (repulsive) scalar potential for fermions is also
attractive (repulsive) for antifermions. For $V_{v}=V_{s}$ and an attractive
vector potential for fermions, the scalar potential is counterbalanced by
the vector potential for antifermions as long as the scalar potential is
attractive and the vector potential is repulsive. As a consequence there is
no bounded solution for antifermions. For $V_{v}=0$ and a pure scalar
attractive potential, one finds energy levels for fermions and antifermions
arranged symmetrically about $E=0$ (see, \textit{e.g.}, Refs. \cite{cn} and
\cite{cnt}). For $V_{v}=-V_{s}$ and a repulsive vector potential for
fermions, the scalar and the vector potentials are attractive for
antifermions but their effects are counterbalanced for fermions. Thus,
recurring to this simple standpoint one can anticipate in the mind that
there is no bound-state solution for fermions in this last case of mixing.

Introducing the unitary operator

\begin{equation}
U(\theta )=\exp \left( -i\frac{\theta }{2}\sigma _{1}\right)  \label{2}
\end{equation}

\noindent where $\theta $ is a real quantity such that $-\pi \leq \theta
\leq \pi $, the transform of the Hamiltonian (\ref{1b}), $H=U\mathcal{H}%
U^{-1}$, takes the form

\begin{equation}
H=\sigma _{1}cp-\sigma _{2}\sin \left( \theta \right) \left(
mc^{2}+V_{s}\right) +\sigma _{3}\cos \left( \theta \right) \left(
mc^{2}+V_{s}\right) +V_{v}  \label{3}
\end{equation}

\noindent In terms of the upper ($\phi $) and the lower ($\chi $) components
of the transform of the spinor $\Psi $ under the action of the operator $U$,
$\psi =U\Psi $, one has
\begin{eqnarray}
\phi &=&\Psi _{+}\cos \left( \frac{\theta }{2}\right) -i\Psi _{-}\sin \left(
\frac{\theta }{2}\right)  \nonumber \\
&&  \label{4} \\
\chi &=&\Psi _{-}\cos \left( \frac{\theta }{2}\right) -i\Psi _{+}\sin \left(
\frac{\theta }{2}\right)  \nonumber
\end{eqnarray}

\noindent Now, as can be seen by inspection of (\ref{4}), $\phi $ and $\chi $
have mixed parities for a parity-invariant theory unless $\theta =0$ or $%
\theta =\pm \pi $. In terms of the components of the new spinor, the Dirac
equation becomes
\begin{eqnarray}
\hbar c\phi ^{\prime }+\sin \left( \theta \right) \left( mc^{2}+V_{s}\right)
\phi  &=&i\left[ E+\cos \left( \theta \right) \left( mc^{2}+V_{s}\right)
-V_{v}\right] \chi   \nonumber \\
&&  \label{4a} \\
\hbar c\chi ^{\prime }-\sin \left( \theta \right) \left( mc^{2}+V_{s}\right)
\chi  &=&i\left[ E-\cos \left( \theta \right) \left( mc^{2}+V_{s}\right)
-V_{v}\right] \phi   \nonumber
\end{eqnarray}

\noindent Choosing
\begin{equation}
V_{v}=V_{s}\cos \left( \theta \right)  \label{5}
\end{equation}
\noindent i.e., $|V_{s}|\geq $ $|V_{v}|$, one has
\begin{eqnarray}
\hbar c\phi ^{\prime }+\sin \left( \theta \right) \left( mc^{2}+V_{s}\right)
\phi &=&i\left[ E+\cos \left( \theta \right) mc^{2}\right] \chi  \label{6a}
\\
&&  \nonumber \\
\hbar c\chi ^{\prime }-\sin \left( \theta \right) \left( mc^{2}+V_{s}\right)
\chi &=&i\left[ E-\cos \left( \theta \right) \left( mc^{2}+2V_{s}\right)
\right] \phi  \label{6b}
\end{eqnarray}

\noindent

\noindent Furthermore, using the expression for $\chi $ obtained from (\ref
{6a}), viz.

\begin{equation}
\chi =-i\,\frac{\hbar c\phi ^{\prime }+\sin \left( \theta \right) \left(
mc^{2}+V_{s}\right) \phi }{E+\cos \left( \theta \right) mc^{2}},\quad E\neq
-\cos \left( \theta \right) mc^{2}  \label{7}
\end{equation}

\noindent and inserting it in (\ref{6b}) one arrives at the following
second-order differential equation for $\phi $:
\begin{equation}
-\frac{\hbar ^{2}}{2}\phi ^{\prime \prime }+\left[ \frac{\sin ^{2}\left(
\theta \right) }{2c^{2}}V_{s}^{2}+\frac{mc^{2}+\cos \left( \theta \right) E}{%
c^{2}}V_{s}-\frac{\hbar \sin \left( \theta \right) }{2c}V_{s}^{\prime }-%
\frac{E^{2}-m^{2}c^{4}}{2c^{2}}\right] \phi =0  \label{8}
\end{equation}

\noindent Therefore, the solution of the relativistic problem is mapped into
a Sturm-Liouville problem for the upper component of the Dirac spinor. In
this way one can solve the Dirac problem by recurring to the solution of a
Schr\"{o}dinger-like problem. For the case of a pure scalar potential ($%
\theta =\pm \pi /2$) with $E\neq 0$, it is also possible to write a simple
second-order differential equation for $\chi $, just differing from the
equation for $\phi $ in the sign of the term involving $V_{s}^{\prime }$,
namely,
\begin{eqnarray}
-\frac{\hbar ^{2}}{2}\phi ^{\prime \prime }+\left[ \frac{V_{s}^{2}}{2c^{2}}%
+mV_{s}\mp \frac{\hbar }{2c}V_{s}^{\prime }\right] \phi &=&\frac{%
E^{2}-m^{2}c^{4}}{2c^{2}}\,\phi  \nonumber \\
&&  \label{8a} \\
-\frac{\hbar ^{2}}{2}\chi ^{\prime \prime }+\left[ \frac{V_{s}^{2}}{2c^{2}}%
+mV_{s}\pm \frac{\hbar }{2c}V_{s}^{\prime }\right] \chi &=&\frac{%
E^{2}-m^{2}c^{4}}{2c^{2}}\,\chi  \nonumber
\end{eqnarray}

\noindent This supersymmetric structure of the two-dimensional Dirac
equation with a pure scalar potential have already been appreciated in the
literature \cite{coo}-\cite{nog}. One can check that the Dirac energy levels
are symmetrical about $E=0$ for a pure scalar potential. This conclusion can
be obtained directly from (\ref{8a}) as well as from the charge conjugation.
Indeed, if $\Psi $ is a solution with energy $E$ then $\sigma _{1}\Psi ^{*}$
is also a solution with energy $-E$ for the very same potential. It means
that the scalar potential couples to the positive-energy component of the
spinor in the same way it couples to the negative-energy component. In other
words, this sort of potential couples to the mass of the fermion instead of
its charge. When a vector potential is present the potentials must undergo
the transformations $V_{v}\rightarrow -V_{v}$ and $V_{s}\rightarrow V_{s}$
under the charge conjugation operation, in order to restore the invariance
of the theory. In addition to the complex conjugate, the upper and lower
components of the Dirac spinor are exchanged by charge conjugation but the
position probability density is invariant (see, \textit{e.g.}, \cite{gre}).

The solutions for $E=-\cos \left( \theta \right) mc^{2}$, excluded from the
Sturm-Liouville problem, can be obtained directly from the Dirac equation (%
\ref{6a})-(\ref{6b}). One can observe that only a pure scalar potential ($%
\theta =\pm \pi /2$, $E=0$) might support such a sort of isolated
normalizable solutions, with the upper and lower components of the Dirac
spinor given by

\begin{eqnarray}
\phi &=&N_{\phi }\exp \left\{ -\frac{\sin \left( \theta \right) }{\hbar c}%
\left[ mc^{2}x+v(x)\right] \right\}  \nonumber \\
&&  \label{9} \\
\chi &=&N_{\chi }\exp \left\{ +\frac{\sin \left( \theta \right) }{\hbar c}%
\left[ mc^{2}x+v(x)\right] \right\}  \nonumber
\end{eqnarray}

\noindent where $N_{\phi }$ and $N_{\chi }$ are normalization constants and $%
v(x)=\int^{x}V_{s}(y)\,dy$. \noindent One can check that it is impossible to
have both components different from zero simultaneously on the same side of
the $x$-axis and that $|\phi (\pm x)|=|\chi (\mp x)|$. Furthermore, $\phi $
and $\chi $ change their roles under the substitution $\theta \rightarrow
-\theta $. Of course a normalizable zero-mode eigenstate is possible only if
$v(x)$ has a distinctive leading asymptotic behaviour.

\section{The inversely linear potential}

Now let us focus our attention on a scalar power-law potential in the form
\begin{equation}
V_{s}=-\frac{\hbar cq}{|x|}  \label{12}
\end{equation}
\noindent where the coupling constant, $q$, is a dimensionless real
parameter.

We begin with the zero-eigenmode solutions. From (\ref{9}) one sees that the
normalizable zero-energy solutions are accomplished only for $q>0$ and they
are expressed by
\begin{equation}
\psi =|x|^{q}\left[ \Theta (-x)S_{-}\exp \left( +\frac{mc}{\hbar }x\right)
+\Theta (+x)S_{+}\exp \left( -\frac{mc}{\hbar }x\right) \right]  \label{12-1}
\end{equation}
where $\Theta (x)$ is the Heaviside step function multiplied by a
normalization constant, and the spinors $S_{\pm }$ are defined by
\begin{equation}
S_{\pm }=\left(
\begin{array}{l}
1 \\
0
\end{array}
\right) ,\qquad S_{\mp }=\left(
\begin{array}{l}
0 \\
1
\end{array}
\right) ,\quad \mathrm{for\quad }\theta =\pm \frac{\pi }{2}  \label{12-2}
\end{equation}

\noindent It follows that the position probability density of the zero-mode
spinor has a lonely hump on each side of the $x$-axis. Furthermore, $q\geq 1$
for obtaining a differentiable spinor at the origin and it means that the
scalar inversely linear potential must be  strong enough  to hold a
zero-mode solution. The finding of a zero-mode solution for a scalar
potential with the same limit for $x\rightarrow +\infty $ and $x\rightarrow
-\infty $ contradicts the statements made in Ref. \cite{nog}. It is
intriguing to find Dirac eigenspinors with a vanishing lower component in a
theory without a nonrelativistic limit. More surprising is to find a
vanishing upper component. Both dramatic circumstances make their appearance
due to the particular form assumed by the upper and the lower components of
the transform of the spinor $\Psi $ given by (\ref{4}). In the presence of a
pure scalar potential Eq. (\ref{4}) gives $\phi =\left( \Psi _{+}-i\Psi
_{-}\right) /\sqrt{2}$ and $\chi =-i\left( \Psi _{+}+i\Psi _{-}\right) /%
\sqrt{2}$. Therefore, in the nonrelativistic regime one obtains $|\phi
|\approx |\chi |$. On the other side, in the ultrarelativistic regime one
expects that $\Psi _{-}$ presents a contribution comparable to $\Psi _{+}$
for nonnegative energies, thus the possibilities $\Psi _{+}\approx i\Psi _{-}
$ and $\Psi _{+}\approx -i\Psi _{-}$ imply into $\phi \approx 0$ and $\chi
\approx 0$, respectively. Therefore, one can conclude that the zero-mode
solutions given by (\ref{12-1}) correspond to the ultrarelativistic limit of
the theory.

We still need to consider the more general case of solutions, i.e., these
ones with $E\neq -\cos \left( \theta \right) mc^{2}$. In this case Eq. (\ref
{8}) becomes

\begin{equation}
-\frac{\hbar ^{2}}{2}\,\phi _{\varepsilon }^{\prime \prime
}+V_{eff}^{\varepsilon }\,\phi _{\varepsilon }=E_{eff}\,\phi _{\varepsilon }
\label{12a}
\end{equation}
where
\begin{equation}
E_{eff}=\frac{E^{2}-m^{2}c^{4}}{2c^{2}}  \label{12b}
\end{equation}
$\varepsilon $ stands for the sign function ($\varepsilon =x/|x|$, for $%
x\neq 0$), and the effective potential is the Kratzer-like potential
\begin{equation}
V_{eff}^{\varepsilon }=-\frac{\hbar cq_{eff}}{|x|}+\frac{A^{\varepsilon }}{%
x^{2}}  \label{13}
\end{equation}
\noindent with
\begin{equation}
q_{eff}=q\,\frac{mc^{2}+\cos \left( \theta \right) E}{c^{2}},\quad
A^{\varepsilon }=\frac{\hbar ^{2}}{2}\,\xi \left( \xi -\varepsilon \right)
,\quad \xi =q\sin \left( \theta \right)   \label{14}
\end{equation}

\noindent Here, such as for the isolated solutions given by (\ref{12-1}),
the space is split into two regions, and $\phi _{+}$ refers to $\phi (x>0)$
and $\phi _{-}$ to $\phi (x<0)$. These last results tell us that the
solution for this class of problem consists in searching for bounded
solutions for two Schr\"{o}dinger equations. Therefore, one has to search
for bound-state solutions for $V_{eff}^{+}$ and $V_{eff}^{-}$ with a common
effective eigenvalue. The Dirac eigenvalues are obtained by inserting the
effective eigenvalues in (\ref{12b}).

Before proceeding, it is useful to make some qualitative arguments regarding
the Kratzer-like potential and its possible solutions. The parameters of the
effective Kratzer-like potential are related in such a manner that the
change $\theta \rightarrow -\theta $ induces the change $V_{eff}^{\pm
}\rightarrow V_{eff}^{\mp }$ ($A^{\pm }\rightarrow A^{\mp }$), meaning that
the effective potential for $\phi _{\pm }$ transforms into the potential for
$\phi _{\mp }$. The effective Kratzer-like potential is able to bind
fermions on the condition that $q_{eff}>0$. It follows that $E_{eff}<0$,
corresponding to Dirac eigenvalues in the range $-mc^{2}<E<+mc^{2}$, and $q>0
$. The energies belonging to $|E|\geq mc^{2}$ correspond to the continuum.
One can see that $\phi _{\varepsilon }$ is subject to a potential-well
structure for $V_{eff}^{\varepsilon }$ when $|\xi |>1$. For $0<|\xi |\leq 1$
the effective potential has a potential-well structure on one of the sides
of the $x$-axis and a singular potential at the origin, with singularity
given by $-1/|x|$ when $|\xi |=1$ and $-1/x^{2}$ when $0<|\xi |<1$, on the
other side of the $x$-axis. For $\xi =0$ the singularity $-1/|x|$ appears on
both sides of the $x$-axis. It is worthwhile to note at this point that the
singularity $-1/x^{2}$ never exposes the fermion to collapse to the center
\cite{lan} because in any condition $A^{\varepsilon }$ is never less than
the critical value $A_{c}=-\hbar ^{2}/8$. The Schr\"{o}dinger equation with
the Kratzer-like potential is an exactly solvable problem and its solution,
for a repulsive inverse-square term in the potential ($A^{\varepsilon }>0$),
can be found on textbooks \cite{lan}-\cite{flu}. Since we need solutions
involving a repulsive as well as an attractive inverse-square term in the
potential, the calculation including this generalization is presented.

Defining the quantities $z$ and $B$,

\negthinspace
\begin{equation}
z=\frac{2}{\hbar }\sqrt{-2E_{eff}}\;|x|,\quad B=q_{eff}\,c\,\sqrt{-\frac{1}{%
2E_{eff}}}  \label{15}
\end{equation}

\noindent and using (\ref{8})-(\ref{9}) and (\ref{13}) one obtains the
equation

\begin{equation}
\,\phi _{\varepsilon }^{\prime \prime }+\left( -\frac{1}{4}+\frac{B}{z}-%
\frac{2A^{\varepsilon }}{\hbar ^{2}z^{2}}\right) \phi _{\varepsilon }=0
\label{16}
\end{equation}

\noindent Now the prime denotes differentiation with respect to $z$. The
normalizable asymptotic form of the solution as $z\rightarrow \infty $ is $%
e^{-z/2}$. As $z\rightarrow 0$, when the term $1/z^{2}$ dominates, the
solution behaves as $z^{s_{\varepsilon }}$, where $s_{\varepsilon }$ is a
solution of the algebraic equation

\begin{equation}
s_{\varepsilon }(s_{\varepsilon }-1)-\frac{2A^{\varepsilon }}{\hbar ^{2}}=0
\label{17}
\end{equation}
viz.

\begin{equation}
s_{\varepsilon }=\frac{1}{2}\left( 1\pm \sqrt{1+\frac{8A^{\varepsilon }}{%
\hbar ^{2}}}\right) =\pm \xi +\frac{1\mp \varepsilon }{2}\geq 0  \label{18}
\end{equation}

\noindent The solution for all $z$ can be expressed as $\phi _{\varepsilon
}(z)=z^{s_{\varepsilon }}e^{-z/2}w(z)$, where $w$ is solution of Kummer\'{}s
equation \cite{abr}

\begin{equation}
zw_{\varepsilon }^{\prime \prime }+(b_{\varepsilon }-z)w_{\varepsilon
}^{\prime }-a_{\varepsilon }w_{\varepsilon }=0  \label{19}
\end{equation}

\noindent with

\begin{equation}
a_{\varepsilon }=s_{\varepsilon }-B,\quad b_{\varepsilon }=2s_{\varepsilon }
\label{20}
\end{equation}

\noindent Then $w_{\varepsilon }$ is expressed as $M(a_{\varepsilon
},b_{\varepsilon },z)$ and in order to furnish normalizable $\phi
_{\varepsilon }$, the confluent hypergeometric function must be a
polynomial. This demands that $a_{\varepsilon }=-n_{\varepsilon }$, where $%
n_{\varepsilon }$ is a nonnegative integer in such a way that $%
M(a_{\varepsilon },b_{\varepsilon },z)$ is proportional to the associated
Laguerre polynomial $L_{n_{\varepsilon }}^{b_{\varepsilon }-1}(z)$, a
polynomial of degree $n_{\varepsilon }$. This requirement, combined with the
first equation of (\ref{20}), also implies into quantized effective
eigenvalues:

\begin{equation}
E_{eff}=-\,\frac{q_{eff}^{2}c^{2}}{2\left( s_{\varepsilon }+n_{\varepsilon
}\right) ^{2}},\qquad n_{\varepsilon }=0,1,2,\ldots  \label{21}
\end{equation}

\noindent with eigenfunctions given by

\begin{equation}
\phi _{\varepsilon }(z)=N_{\phi _{\varepsilon }}\;z^{s_{\varepsilon
}}e_{\;}^{-z/2}\;L_{n_{\varepsilon }}^{2s_{\varepsilon }-1}\left( z\right)
,\qquad s_{\varepsilon }>0  \label{22}
\end{equation}

\smallskip

\noindent

\noindent $N_{\phi _{\varepsilon }}$ is a normalization constant and the
constraint over $s_{\varepsilon }$ is a consequence of the definition of
associated Laguerre polynomials which is going to imply that $q>0$, as
advertized by the preceding qualitative arguments. If $A^{\varepsilon }>0$
there is just one possible value for $s_{\varepsilon }$  and the same is
true for $A^{\varepsilon }=A_{c}$ when $s_{\varepsilon }=1/2$, but for $%
A_{c}<A^{\varepsilon }<0$ there are two possible values for $s_{\varepsilon }
$ in the interval $0<s<1$. If the inverse-square potential is absent ($%
A^{\varepsilon }=0$) then $s_{\varepsilon }=1$. Note that the behavior of $%
\phi _{\varepsilon }$ at very small $z$ implies into the Dirichlet boundary
condition ($\phi _{\varepsilon }(0)=0$). This boundary condition is
essential whenever $A^{\varepsilon }\neq 0$, nevertheless it also develops
for $A^{\varepsilon }=0$.

The necessary conditions for binding fermions in the Dirac equation with the
effective Kratzer-like potential have been put forward. The formal
analytical solutions have also been obtained. Now we move on to consider a
survey for distinct cases in order to match the common effective eigenvalue
on both sides of the $x$-axis. As we will see this survey leads to
additional restrictions on the solutions, including constraints involving
the nodal structure of the Dirac spinor.

For\textbf{\ }$\sin \left( \theta \right) <0$ one has $n_{-}=n+1$, where $%
n=n_{+}$ ($s_{-}=s_{+}-1$). For\textbf{\ }$\sin \left( \theta \right) =0$
one has $n_{-}=n$ ($s_{-}=s_{+}$), and for $\sin \left( \theta \right) >0$
one has $n_{-}=n-1$ ($s_{-}=s_{+}+1$). The Dirac eigenvalues can now be
written as

\begin{equation}
E=mc^{2}\,\frac{-\left( \frac{q}{|\xi |+n}\right) ^{2}\cos \left( \theta
\right) \pm \sqrt{1-\left( \frac{\xi }{|\xi |+n}\right) ^{2}}}{1+\left(
\frac{q\cos \left( \theta \right) }{|\xi |+n}\right) ^{2}}  \label{23b}
\end{equation}

\noindent and the upper component of the Dirac spinor on the positive
half-line is given by

\begin{equation}
\phi =Ne_{\;}^{-z/2}\left\{
\begin{array}{c}
zL_{n}^{1}\left( z\right) \\
\\
z^{|\xi |}L_{n}^{2|\xi |-1}\left( z\right)
\end{array}
\begin{array}{l}
,\;\textrm{for}\;\xi =0 \\
\\
,\;\textrm{for}\;\xi >0
\end{array}
\right. \;  \label{23a}
\end{equation}

\noindent where, as before, $\xi =q\sin \left( \theta \right) $. $N$ is a
normalization constant and $n=1,2,3,\ldots $ ($n=0$ is to be included for
considering the zero-eigenmodes of a pure scalar potential ($\theta =\pm \pi
/2$) in the event that $q\geq 1$). We have used $L_{-1}^{k}\left( z\right)
=0 $ for all $k$. For $n=0$ the solution for $\chi $ is already embraced in (%
\ref{12-1}) and for $n\neq 0$ it can be obtained by using (\ref{7}). By
using some recurrence relations involving the associated Laguerre
polynomials \cite{abr}, one can find that

\begin{equation}
\chi =\mp iq\,\frac{n+1}{n}\,Ne_{\;}^{-z/2}\left[
L_{n}^{0}(z)+L_{n+1}^{0}(z)\right] ,\left\{
\begin{array}{l}
-\;\textrm{for}\;\theta =0 \\
+\;\textrm{for}\;\theta =\pm \pi
\end{array}
\right.  \label{31a}
\end{equation}
\[
\chi =-\frac{i}{\xi }\,\frac{q}{\xi +n}\frac{E\cos (\theta )+mc^{2}}{%
E+mc^{2}\cos (\theta )}Nz^{|\xi |}e_{\;}^{-z/2}\qquad \qquad \qquad \qquad
\qquad \qquad \qquad \qquad
\]

\begin{equation}
\times \left\{ \left[ \frac{mc^{2}}{E\cos (\theta )+mc^{2}}\frac{\xi +n}{q}%
\sin (\theta )-1\right] L_{n}^{2|\xi |-1}\left( z\right)
-2L_{n-1}^{2|\xi |}\left( z\right) \right\} ,\;\textrm{for}\;\xi
>0  \label{31b}
\end{equation}

\noindent It follows from (\ref{23a}) and (\ref{31b}) that for $\theta =\pi
/2$, $\phi $ and $\chi $ are proportional to $z^{q}e_{\;}^{-z/2}L_{n}^{2q-1}%
\left( z\right) $ and $z^{q+1}e_{\;}^{-z/2}L_{n-1}^{2q+1}\left( z\right) $,
respectively, in agreement with the result found in Ref. \cite{asc6}. The
upper and lower components of the Dirac spinor for $\xi <0$ can be obtained
from those ones for $\xi >0$ by changing $\xi $ by $\xi +1$ and $n$ by $n-1$
in $\phi $, and $n$ by $n+1$ in $\chi $. It is worthwhile to note that $\chi
$ satisfies the Dirichlet boundary condition, an exception is for $V_{v}=\pm
V_{s}$, a circumstance when $\chi $ is proportional to the first derivative
of $\phi $. Anyway, $|\chi (0)|<<1$ when $q<<1$ in such a manner that the
original spinor $\Psi $ has a lower (upper) component suppressed relative to
the upper (lower) component for $V_{v}=V_{s}$ ($V_{v}=-V_{s}$), as can be
seen from Eq. (\ref{4}).

A differentiable spinor at the origin is always possible for $V_{v}=\pm V_{s}
$, but for $V_{v}\neq \pm V_{s}$ an acceptable solution at the origin can be
achieved only if $|\xi |>1$, i.e., $q\geq 1/|\sin (\theta )|$ (remember that
such a restriction on the coupling constant has already appeared in the case
of the zero-eigenmode for a pure scalar potential). Therefore, an
appropriate nonrelativistic limit of the theory becomes possible only if $%
V_{v}=V_{s}$ because only in this case one can consider a weakly attractive
potential for fermions. The Dirac eigenenergies given by (\ref{23b}) are
invariant under the substitution $\theta \rightarrow -\theta $. One can say
that this happens because this transformation does not alter the mixing
among the vector and scalar potentials. Nevertheless, $\phi $ and $\chi $
are affected by the presence of $\sin (\theta )$ into the Dirac equation. We
have already seen its effects for the zero-eigenmode spinor and that for $%
E\neq -\cos (\theta )mc^{2}$ it implies into $V_{eff}^{\pm }\rightarrow
V_{eff}^{\mp }$. Inspection of (\ref{4a}) reveals that $\phi _{\pm
}\rightarrow \phi _{\mp }^{*}$ and $\chi _{\pm }\rightarrow \chi _{\mp }^{*}$%
. Therefore, one can conclude that the change $\theta \rightarrow -\theta $
does not alter the state of a fermion because it just changes $\psi (x)$ by $%
\psi ^{*}(-x)$ while maintains its eigenenergy. Furthermore, for $V_{v}=0$
the transformation $\theta \rightarrow -\theta $ exchanges the upper and
lower components of the Dirac spinor. The results for $\theta =\pm \pi
/2+\delta $, where $-\pi /2<\delta <\pi /2$, can be obtained from the
results for $\theta =\pm \pi /2-\delta $ by changing $E$ by $-E$.

When $\theta =\pm \pi /2$, the case of a pure scalar potential, the energy
levels are given by
\begin{equation}
E=\pm mc^{2}\,\sqrt{1-\left( \frac{q}{q+n}\right) ^{2}},\qquad n=0,1,2,\ldots
\label{23d}
\end{equation}

\noindent so that the energy levels for fermions and antifermions are
symmetric about $E=0$. Note that $E\approx mc^{2}$ only if $n\gg q$, thus
the nonrelativistic limit of the theory would be, in a limited sense, a
regime of large quantum numbers. On the other hand, in the regime of strong
coupling, \textit{i.e.}, for $q\gg 1$, one has $E\approx mc^{2}n/q$ and as
the coupling becomes extremely strong the lowest effective eigenvalues end
up close to zero. Now one sees clearly that the eigenvalues for a
zero-energy solution, in contrast to what is declared in Ref. \cite{cnt},
can be obtained as a limit case of a nonzero-energy solution. The Dirac
eigenenergies are plotted in Fig. \ref{Fig1} for the four lowest bound
states as a function of $\theta =\pi /2+\delta $. Starting from $\pi /2$, as
$\theta $ is increased ($\delta >0$) all the energy levels move toward the
upper continuum. On the other side, as $\theta $ decreases ($\delta <0$) all
the energy levels move toward the lower continuum. The mixed vector-scalar
Coulomb potential present a continuous transition by starting from $\theta
=\pm \pi /2$, with energy levels for fermions and antifermions always
present. When $\theta =0$ or $\theta =\pm \pi $, though, there is a clear
phase transition. The phase transition shows its face not only for the
energy levels but also for the eigenspinor. This phenomenon  is due to the
abrupt disappearance of the singularity $1/x^{2}$ in the effective
potential. For $\theta =0$ ($V_{v}=V_{s}$) the energy levels given by

\begin{equation}
E=mc^{2}\,\frac{n^{2}-q^{2}}{n^{2}+q^{2}},\qquad n=1,2,3,\ldots  \label{23c}
\end{equation}

\noindent are pushed down from the upper continuum so that these energy
levels correspond to bound states of fermions (see Fig. \ref{Fig2}). In this
case there are no energy levels for antifermions. All the Dirac eigenvalues
are positive if $q<1$, and some negative eigenvalues arise if $q>1$. One has
$E-mc^{2}\approx -2q^{2}/n^{2}$ as long as $q\ll n$. When $\theta =\pm \pi $
only the energy levels emerging from the lower continuum, the energy levels
for antifermions, survive:
\begin{equation}
E=-mc^{2}\,\frac{n^{2}-q^{2}}{n^{2}+q^{2}},\qquad n=1,2,3,\ldots  \label{23e}
\end{equation}

\noindent For this case the energy levels are illustrated in Fig. \ref{Fig3}
as a function of $q$. Note that $E\approx mc^{2}$ only in the
strong-coupling regime.

In all the circumstances, namely $|V_{s}|\geq |V_{v}|$, there is no
atmosphere for the spontaneous production of particle-antiparticle pairs. No
matter the signs of the potentials or how strong they are, the positive- and
negative-energy levels neither meet nor dive into the continuum. Thus there
is no room for the production of fermion-antifermion pairs. This all means
that Klein\'{}s paradox never comes to the scenario.

Figs. \ref{Fig4}, \ref{Fig5} and \ref{Fig6} illustrate the behavior of the
upper and lower components of the Dirac spinor, $|\phi |^{2}$ and $|\chi
|^{2}$, and the position probability density, $|\psi |^{2}=|\phi |^{2}+|\chi
|^{2}$, on the positive side of the $x$-axis for the positive-energy
solutions, with $n=1$, for $\theta =0,\pi /4$ and $\pi /2$, respectively.
The normalization constant was obtained by numerical computation. Since the
inversely linear potential given by (\ref{12}) is invariant under reflection
through the origin ($x\rightarrow -x$), eigenfunctions of the original
Hamiltonian given by (\ref{1b}) with well-defined parities can be found. For
$\theta \neq 0,\pm \pi $, those eigenfunctions can be constructed by taking
symmetric and antisymmetric linear combinations of $\Psi _{+}$ and $\Psi _{-}
$. These new eigenfunctions are continuous everywhere and possess the same
Dirac eigenvalue, then there is a two-fold degeneracy. Nevertheless, the
matter is a little more complicated for $\theta =0,\pm \pi $. For $\theta
\neq 0,\pm \pi $, the effective potential for $\phi $ always presents a
positive singularity so that it makes sense to consider only the half-line.
For $\theta =0,\pm \pi $, though,   there are attractive singularities on
both sides of the $x$-axis, so that the behaviour of a fermion on one side
of the $x$-axis is sensitive to what happens on the other side. Therefore,
the entire line has to be considered. In these last circumstances $\Psi
_{+}=\phi $ and $\Psi _{-}=\chi $, for $\theta =0$, and $\Psi _{+}=\pm i\chi
$ and $\Psi _{-}=\pm i\phi $, for $\theta =\pm \pi $. Recall that $\phi $
vanishes at the origin but $\chi $ does not, so one of those symmetric and
antisymmetric linear combinations of $\Psi _{+}$ and $\Psi _{-}$ is
discontinuous at the origin. In fact, the pair of first-order differential
equations given by (\ref{4a}) implies that $\Psi _{+}$ and $\Psi _{-}$ can
be discontinuous wherever the potential undergoes an infinite jump. In the
specific case under consideration, the effect of the singularity of the
potential can be evaluated by integrating (\ref{4a}) from $-\delta $ to $%
+\delta $ and taking the limit $\delta \rightarrow 0$. Since $\Psi _{+}$ and
$\Psi _{-}$ have opposite parities, the connection conditions can be
summarized in the couple of formula:

\begin{equation}
\left.
\begin{array}{c}
\Psi _{-}(+\delta )=iq\int_{-\delta }^{+\delta }dx\;\frac{\Psi _{+}}{|x|} \\
\\
\int_{-\delta }^{+\delta }dx\;\Psi _{-}=0
\end{array}
\begin{array}{l}
,\;\textrm{for}\;\Psi _{+}\;\textrm{even} \\
\\
,\;\textrm{for}\;\Psi _{+}\;\textrm{odd}
\end{array}
\right.  \label{24}
\end{equation}

\noindent One can verify that the first connection condition is not
satisfied for $\theta =0$, while the second one is not satisfied for $\theta
=\pm \pi $. Therefore, we are forced to  conclude that the upper component
of the original Dirac spinor ($\Psi _{+}$) must be an odd-parity function
for $V_{v}=V_{s}$, and an even-parity function for $V_{v}=-V_{s}$, so that
the bound-state solutions for $V_{v}=\pm |V_{s}|$ are nondegenerate.

\section{Conclusions}

We have succeed in searching for exact Dirac bounded solutions for massive
fermions by considering a convenient mixing of vector-scalar inversely
linear potentials in 1+1 dimensions. The satisfactory completion of this
task has been alleviated by the methodology of effective potentials which
has transmuted the question into Schr\"{o}dinger-like equations with
effective Kratzer-like potentials.

Isolated solutions have also been searched and they have been found only in
the special case of a pure scalar potential. We have shown that those
isolated zero-energy solutions are consistent with the ultrarelativistic
limit of the theory. The existence of such zero-eigenmodes does not conform
with the ``topological'' criterion of Ref. \cite{nog}, which requires that
the scalar potential has different limits for $x\rightarrow +\infty $ and $%
x\rightarrow -\infty $. From Eq. (\ref{9}) one can see that there can be
other sorts of scalar potentials holding zero-energy solutions which do not
satisfy the ``topological'' criterion. Such a criterion is circumvented
because the upper and lower components of the Dirac spinor are normalizable
on the entire line, although they are not simultaneously normalizable on
each side of the $x$-axis, because they can not be simultaneously different
from zero there.

\begin{figure}[th]
\begin{center}
\includegraphics[width=9cm, angle=270]{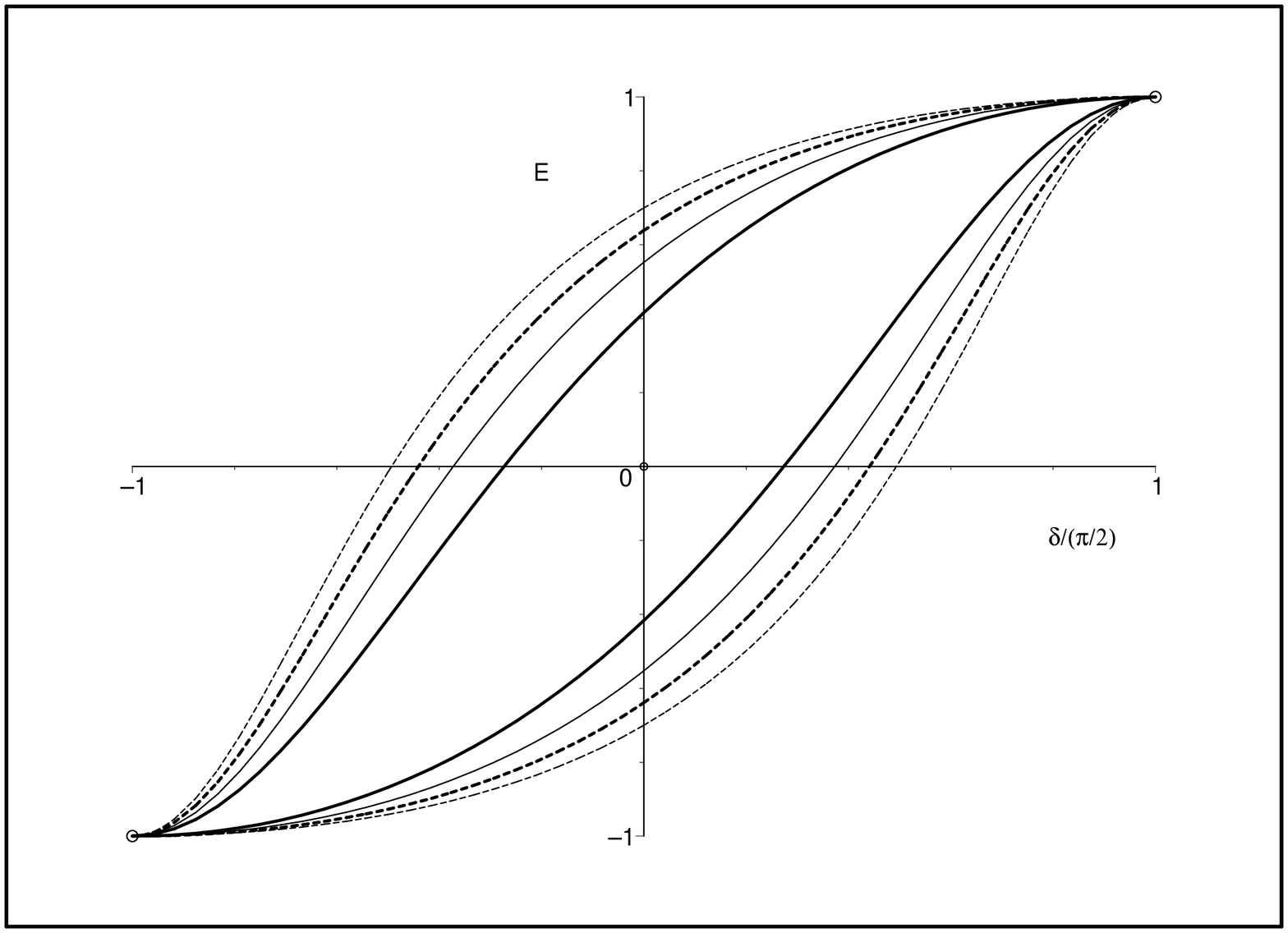}
\end{center}
\par
\vspace*{-0.1cm}
\caption{Dirac eigenvalues for the four lowest energy levels as a function
of $\delta $ ($V_{v}=-V_{s}\sin(\delta )$, with $-\pi /2<\delta <\pi /2$).
The full thick line stands for $n=1$, the full thin line for $n=2$, the
heavy dashed line for $n=3$ and the light dashed line for $n=4$ ($m=c=1$ and
$q=10/|\sin(\theta )|$). The isolated point in $\delta =0$ ($E=0$) is always
present and it corresponds to the zero-eigenmode with $n=0$.}
\label{Fig1}
\end{figure}

For $-|V_{s}|<V_{v}<+|V_{s}|$, there exist bound-state solutions for
fermions and antifermions and the plot of the eigenenergy as a function of
the mixing parameter, $\delta $, looks like a hysteresis loop (Fig. \ref
{Fig1}). Those two-fold degenerate bounded solutions do not present a
nonrelativistic limit because the coupling constant, $q$, can never be a
small quantity.

\begin{figure}[th]
\begin{center}
\includegraphics[width=9cm, angle=270]{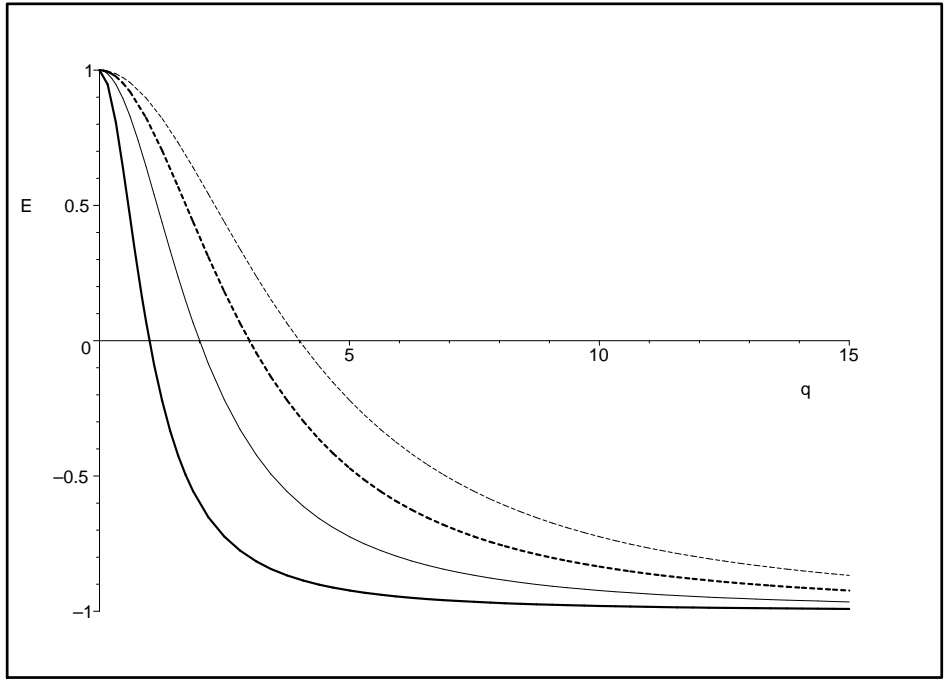}
\end{center}
\par
\vspace*{-0.1cm}
\caption{Dirac eigenvalues for the four lowest energy levels as a function
of $q$ for $\theta = 0$ ($V_{v}=V_{s}$). The full thick line stands for $n=1$%
, the full thin line for $n=2$, the heavy dashed line for $n=3$ and the
light dashed line for $n=4$ ($m=c=1$). }
\label{Fig2}
\end{figure}

\begin{figure}[th]
\begin{center}
\includegraphics[width=9cm, angle=270]{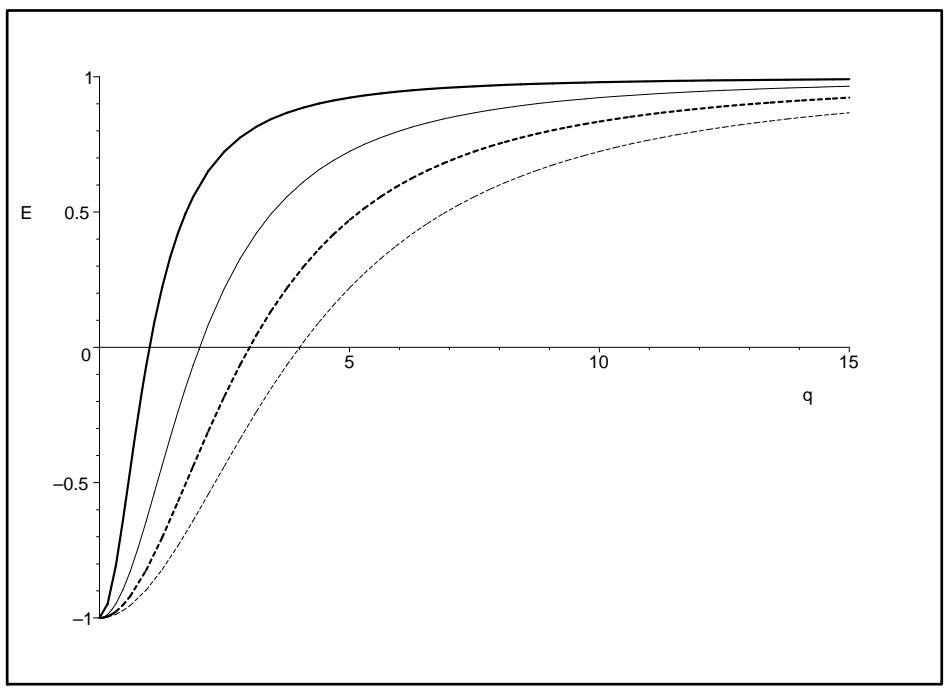}
\end{center}
\par
\vspace*{-0.1cm}
\caption{The same as in Fig. 2 for $\theta = \pi$ ($V_{v}=-V_{s}$). }
\label{Fig3}
\end{figure}

\begin{figure}[th]
\begin{center}
\includegraphics[width=9cm, angle=270]{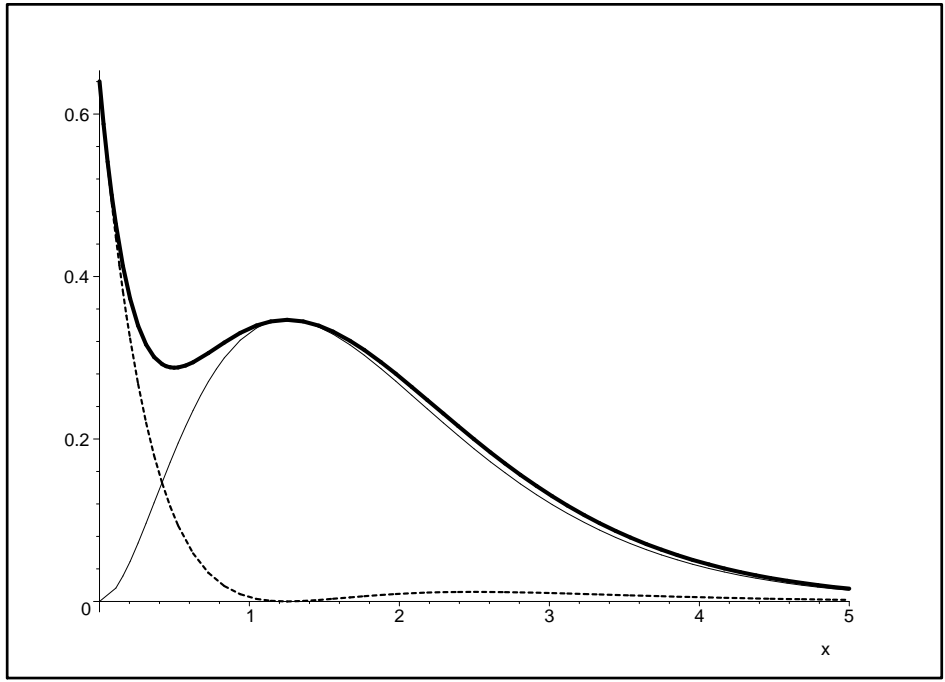}
\end{center}
\par
\vspace*{-0.1cm}
\caption{$|\phi |^{2}$ (full thin line), $|\chi |^{2}$ (dashed line) and $%
|\psi |^{2}=|\phi _{+}|^{2}+|\chi |^{2}$ (full thick line) as a function of $%
x$, corresponding to the ground state ($n=1$) for $\theta =0$ with $q=1/2$ ($%
m=c=\hbar =1$). }
\label{Fig4}
\end{figure}

\begin{figure}[th]
\begin{center}
\includegraphics[width=9cm, angle=270]{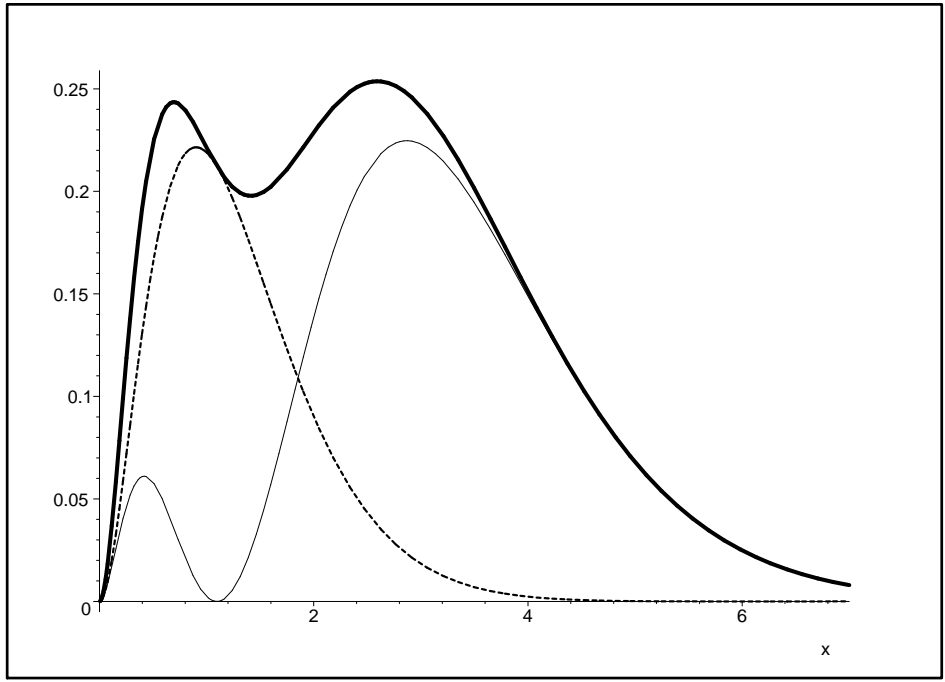}
\end{center}
\par
\vspace*{-0.1cm}
\caption{$|\phi |^{2}$ (full thin line), $|\chi |^{2}$ (dashed line) and $%
|\psi |^{2}=|\phi _{+}|^{2}+|\chi |^{2}$ (full thick line) as a function of $%
x$, corresponding to the positive-ground-state energy ($n=1$) for $\theta
=\pi /4$ with $q=\protect\sqrt{2}$ ($m=c=\hbar =1$). }
\label{Fig5}
\end{figure}

\begin{figure}[th]
\begin{center}
\includegraphics[width=9cm, angle=270]{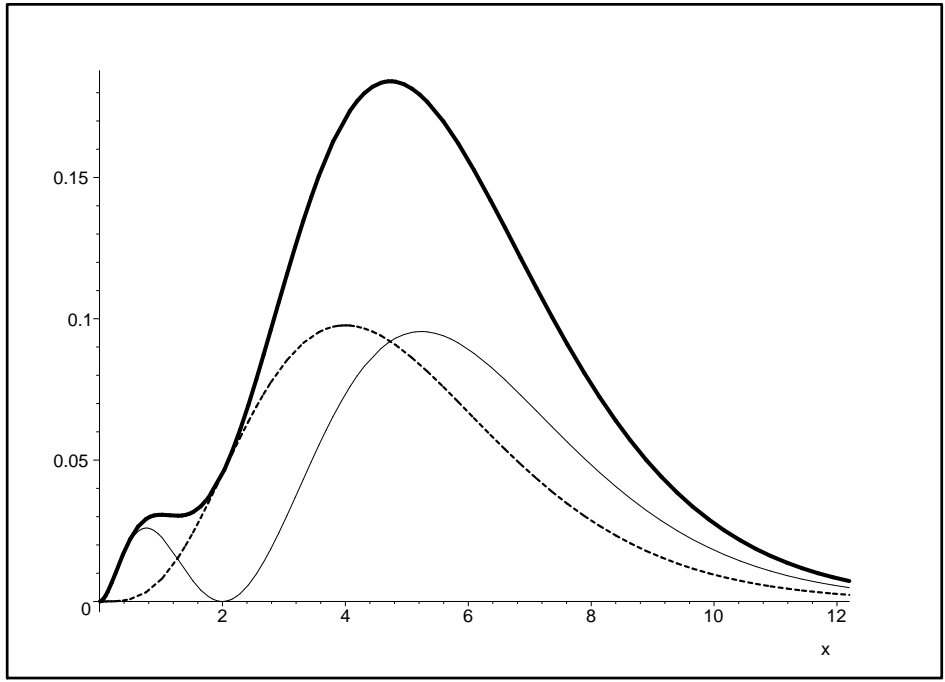}
\end{center}
\par
\vspace*{-0.1cm}
\caption{$|\phi |^{2}$ (full thin line), $|\chi |^{2}$ (dashed line) and $%
|\psi |^{2}=|\phi _{+}|^{2}+|\chi |^{2}$ (full thick line) as a function of $%
x$, corresponding to the positive-first-excited-state energy ($n=1$) for $%
\theta =\pi /2$ with $q=1$ ($m=c=\hbar =1$). }
\label{Fig6}
\end{figure}

For the ``saturation points'', viz. $V_{v}=\pm |V_{s}|$, there are
bound-state solutions either for fermions or for antifermions (Figs. \ref
{Fig2} and \ref{Fig3}). Those phase transitions manifest not only for the
energy levels but also for the eigenspinor as well as for the coupling
constant. Furthermore, the solutions become discontinuous at the origin. A
careful analysis of those discontinuities shows that the potential can only
hold bounded solutions when the upper component of the Dirac spinor behaves
as an odd (even)-parity function for $V_{v}=+|V_{s}|$ ($V_{v}=-|V_{s}|$).
Therefore, the phase transitions transform two-fold degenerate solutions
into nondegenerate ones. In the ``critical points'' the coupling constant
can assume any value and for the special case $V_{v}=+|V_{s}|$ the theory
presents a definite nonrelativistic limit ($q\ll 1$ and $E\simeq mc^{2}$).

Beyond its intrinsic importance as a new solution for a fundamental equation
in physics, the problem analyzed in this paper presents unusual results.
Moreover, it favors the conclusion that even-parity solutions to the
nonrelativistic one-dimensional hydrogen atom do not exist.

\bigskip\bigskip\bigskip\bigskip

\noindent{\textbf{Acknowledgments} }

This work was supported in part by means of funds provided by CNPq and
FAPESP.

\medskip \newpage

\end{document}